\documentclass[12pt]{article}
\input epsf.tex
\def\DESepsf(#1 width #2){\epsfxsize=#2 \epsfbox{#1}}
\usepackage{epsfig}
\usepackage{color}
\usepackage{array}
\makeatletter
\def\alt{\mathrel{\mathpalette\gl@align<}}
\def\agt{\mathrel{\mathpalette\gl@align>}}
\def\gl@align#1#2{\lower.6ex\vbox{\baselineskip\z@skip\lineskip\z@
\ialign{$\m@th#1\hfil##\hfil$\crcr#2\crcr\sim\crcr}}}
\makeatother

\begin{document}
\begin{flushright}
{\tt hep-ph/0402113}\\
February, 2004 \\
UMD-PP-04-019\\
\end{flushright}
\vspace*{2cm}
\begin{center}
{\baselineskip 25pt \large{\bf CKM CP Violation in a Minimal SO(10) Model
for Neutrinos and Its Implications
} \\
}

\vspace{1cm}

{\large
Bhaskar Dutta{$^*$},
Yukihiro Mimura{$^*$}
and R.N. Mohapatra{$^\dagger$}

\vspace{.5cm}

{\it 
$^*$Department of Physics, University of Regina, \\
Regina, Saskatchewan S4S 0A2, Canada \\
$^\dagger$Department of Physics, University of Maryland, \\
College Park, MD 20742, USA\\
}}
\vspace{.5cm}

\vspace{1.5cm}
{\bf Abstract}
\end{center}

A minimal supersymmetric SO(10) model
with one {\bf 10} and one {\bf 126} Higgs superfield has
recently been shown to predict all neutrino mixings as well as the solar
mass difference squared in agreement with observations. Two assumptions
critical to the predictivity and success of the model are
that: (i) the superpotential includes only renormalizable terms, thereby
limiting the number of free parameters 
and (ii) the triplet term in the type II seesaw formula for neutrino mass
dominates, leading to the sum rule $M_\nu~=~c(M_d-M_e)$ that is
responsible for large mixings. However, CKM CP phase is constrained to be
in the second or third quadrant requiring a significant non-CKM component
to CP
violation to explain observations.  We revisit this issue using type I seesaw formula for
neutrino masses and obtain the following results: (i) we show that the
above sumrule
responsible for large mixing angles can also emerge in type I seesaw
models; the detailed
predictions are however not compatible with present data for any
choice of CP phases. (ii) We then show that addition of a
nonrenormalizable term restores compatibility with neutrino data and CKM
CP violation both in type I and type II cases.  We further find that
(iii) the MSSM parameter $\tan\beta\geq 30$ in the type I model and 
 (iv) lepton flavor violation and lepton electric
dipole moments which are accessible to proposed experiments
in both type I and type II models. We also discuss the unification of the gauge
couplings in type I model which requires an intermediate scale.

\thispagestyle{empty}

\bigskip
\newpage

\addtocounter{page}{-1}

\section{Introduction}
\baselineskip 20pt

The observations involving the solar and atmospheric neutrinos
together with those using the accelerator and reactor neutrinos
have now conclusively established that
neutrinos have mass and they mix among themselves.
In conjunction with the negative results from CHOOZ
and PALO-VERDE reactor experiments, a reasonably clear outline of the
mixing pattern among the three generations of neutrinos has emerged.
Of the three angles needed to characterize these mixings,
$\theta_{12}$, $\theta_{23}$ and $\theta_{13}$,
the first two, responsible for solar and atmospheric neutrino deficits
respectively, are large,
and the third corresponding to reactor neutrinos is small
\cite{Bahcall:2002hv, 
Barger:2003qi}. One of the major experimental issues in neutrino physics
now is to make the knowledge of these angles more precise.

Foremost among the theoretical challenges that have already emerged from
these discoveries are, first an  understanding of the smallness of
neutrino masses and second, understanding the vastly different pattern of
mixings among neutrinos from the quarks. Specifically, a key question 
is whether it is possible
to reconcile the large neutrino mixings with small quark mixings in grand
unified frameworks that unify quarks and leptons.

The first challenge, i.e. the lightness of neutrino masses
is elegantly answered by the seesaw mechanism \cite{Mohapatra:1979ia}
which requires an extension of the standard model that includes heavy 
right handed neutrinos. The light neutrino mass matrix is obtained by
integrating out heavy right handed neutrinos and one gets
\begin{equation}
{\cal M}_\nu^{\rm I} = - M_\nu^D M_R^{-1} (M_\nu^D)^T,
\end{equation}
where $M_\nu^D$ is the Dirac neutrino mass matrix and $M_R$ is the 
right handed Majorana mass matrix.
It is therefore important to explore whether one can answer the
second puzzle of large neutrino mixings within the seesaw framework.

The above formula for the neutrino mass matrix is called the type I seesaw
formula. 
The right handed Majorana mass scale, $v_R$, is almost determined 
by the mass squared difference needed to understand the atmospheric
neutrino data to be around $10^{14}$ GeV, if we assume that the Dirac
neutrino mass is same as up-type quark mass. Before proceeding to discuss
the implications of this large right handed neutrino mass, let us discuss
the nature of the seesaw formula.

We will consider a class of models where the right handed neutrino mass
is not put in by hand but arises from a renormalizable coupling of the
form $f N N \Delta_R$, where $N$ is a right handed neutrino, $f$ is a
coupling
constant and $\Delta_R$ is a Higgs field whose vacuum expectation value (vev)
gives mass to the right handed neutrino. This is a natural feature of
models with asymptotic parity conservation, such as those based on
$SU(2)_L\times SU(2)_R \times U(1)_{B-L}$ or any higher gauge group 
such as SO(10)), where the $\Delta_R$ field is part of an
$SU(2)_R$ triplet field. Parity invariance then implies that
 we also have an $f \nu \nu \Delta_L$ coupling term as a parity partner
of the $NN\Delta_R$ coupling. In this class of
theories, whenever $\Delta_R$ acquires a vev, so does $\Delta_L$ and they
are related by the formula 
$\langle\Delta_L \rangle \equiv v_L~=~\frac{v^2_{w}}{\lambda
v_R}$, where $v_{w}$ is a weak scale and $v_R$ is the $\Delta_R$ vev
and $\lambda$ is a coupling constant in the Higgs potential. 
The $\Delta_L$ vev contributes a separate seesaw suppressed
Majorana mass to the neutrino leading to a modified seesaw formula for
neutrino masses given below.
\begin{equation}
{\cal M}_\nu^{\rm II} = M_L - M_\nu^D M_R^{-1} (M_\nu^D)^T,
\end{equation}
where $M_L = f v_L$ and $M_R=f v_R$.
This formula for the neutrino mass matrix is called the type II seesaw formula
\cite{Mohapatra:1980yp}.
In the case where right handed Majorana masses are heavy enough,
%(e.g. around GUT scale), 
the second term in the type II seesaw formula 
can be negligible, and the first term, $M_L~=~f v_L$,
is dominant. We will call this pure type II seesaw. When both terms are
comparable, we will call this mixed type II seesaw.

Coming to the large scale for right handed neutrino mass (i.e. $10^{14}$
GeV or higher), we note that it
suggests supersymmetric grand unified theory (GUT) as a natural framework 
to study neutrino masses, since 
experimental data suggests the gauge coupling unification scale in 
 the minimal supersymmetric extension of  the Standard Model (MSSM)
to be, $M_G \sim 2 \times 10^{16}$ GeV, which is close to the seesaw
scale.
The minimal grand unification model for neutrinos is the one
which is based on the SO(10) group since all standard model fermions and
the right handed neutrino fit into
the $\mathbf{16}$-dimensional representation of SO(10),
resulting not only in a complete unification of the quarks and leptons but
also yielding possible relations between the quark and lepton mass
matrices.
One may therefore hope that the neutrino oscillation parameters might be
predictable in an SO(10) theory.

There are two simple routes to realistic SO(10) model building. In the
first class, one may have smaller representations for the Higgs fields 
like {\bf 10} and {\bf 16} multiplets. In this case, 
the nonrenormalizable terms are  added to the superpotential to implement the seesaw
mechanism. These models have the disadvantage that
they break R-parity which then induces rapid proton decay at an
unacceptable level.

An alternative is to introduce both {\bf 10} and $\overline{\mathbf{126}}$
Higgs
multiplets to give fermion masses. In this class of models, the  R-parity is an automatic
symmetry of the model. This naturally prevents the baryon and lepton number
violating terms that give rise to rapid proton decay and also
guarantees a naturally stable supersymmetric dark matter.

In SO(10) models of this type, both the
$\Delta_L$ and $\Delta_R$ fields are part of
the $\overline{\mathbf{126}}$ multiplet. The above mentioned couplings that
contribute to seesaw formula arise from the
couplings of {\bf 16} matter spinors to 
$\overline{\mathbf{126}}$ multiplet. If the
rest of the Higgs
sector is appropriately chosen, this leads to type II seesaw formula for
neutrinos.

 An interesting class of renormalizable SO(10) models with 
$\overline{\mathbf{126}}$
was proposed in Ref.\cite{Babu:1992ia}
 where it was shown that if
in addition to the Higgs multiplets that break SO(10) and do
not couple to fermions,  one chooses only one $\mathbf{10}$ and one
$\overline{\mathbf{126}}$  Higgs field, then the model provides a
completely realistic description of
the charged fermion sector of the standard model and
is very predictive in the neutrino sector without any extra symmetry 
assumptions. The reason for this is that all the Yukawa parameters of the
model are determined by the quark and charged lepton sector.
This model is called the minimal renormalizable SO(10) model, since
unlike the SO(10) models with {\bf 16} Higgs fields, it does not
add any nonrenormalizable terms to the superpotential. The presence of the
$\overline{\mathbf{126}}$ representation that allows 
the neutrino flavor structure to be related to  other fermion mass
matrices is at the heart of the predictivity of the model. 

In the  Refs.\cite{Lavoura:1993vz,Matsuda:2000zp,Fukuyama:2002ch},
this model was analyzed using the type I seesaw formula to see whether the
neutrino oscillation parameters
predicted would agree with observations.
%They found that the Kobayashi-Maskawa (KM) phase 
%can be in first quadrant in the 
%$\rho$-$\eta$ plane
They found that the atmospheric and solar mixings can be large,
and the phenomenological predictions are studied \cite{Fukuyama:2003hn}. 
However they found
that mass squared ratio for the solar and atmospheric data
is $\Delta m^2_\odot/\Delta m^2_A \simeq 0.2$.
This result is incompatible with the recent combined
data analysis 
of the solar mixing angle and mass squared ratio \cite{Barger:2003qi}.
%In Ref.\cite{Fukuyama:2003hn}, leptonic flavor violation, 
%electric dipole moment (EDM) of electron 
%and neutrino magnetic moment are calculated in the type I seesaw version
%of the minimal renormalizable SO(10) model.

A new approach to discussing neutrinos in this model was presented
during the past year. Considering only the second and third generation
sector, the authors of Ref.\cite{Bajc:2002iw} pointed out that
the type II seesaw in the minimal SO(10) model
can provide a natural way to understand the maximal atmospheric mixing
due to the convergence of bottom quark and tau lepton masses when
extrapolated to GUT scale. The reason for this that in the
the pure-type II seesaw case, the $\overline{\mathbf{126}}$
Higgs coupling leads
the neutrino mass matrix to be proportional to $M_d-M_e$, where $M_d$
is a down-type quark mass matrix and $M_e$ is a charged-lepton mass
matrix.  
It is then easy to see that $b$-$\tau$ mass convergence makes the 3-3
entry of the neutrino mass matrix very small, leading to large atmospheric
mixing angle. While this observation was interesting, a priori, it was not
clear if this would survive once the model is extended to
include three generations. It was however shown in 
Ref.\cite{Goh:2003sy}, that the same $b$-$\tau$ mass
convergence that led to large atmospheric mixing, also leads to
large solar mixing while keeping the 1-3 mixing angle 
small as required by data. It also predicts the ratio of solar to
atmospheric mass squared difference i.e. $\Delta m^2_{\odot}/\Delta
m^2_A$ to be in the right range. This establishes that the
minimal renormalizable SO(10) models with {\bf 126} Higgs provide a very
interesting way to understand neutrino masses and mixings within a
complete quark-lepton unified framework.

In Ref.\cite{Goh:2003sy}, all the CP phases, including the
Kobayashi-Maskawa (KM) phase,
were set to zero (or 180 degree) and it was assumed that all CP violating
effects owe their origin to the SUSY breaking sector.
In Ref.\cite{Goh:2003hf},  the pure type II model including
all CP phases was analyzed, and it was found that one can maintain the
predictivity of the model despite the appearance of the phases; but 
the KM phase must be in the second quadrant in the $\rho$-$\eta$ plane
to maintain the neutrino predictions (where we have adopted the
Wolfenstein parameterization for the CKM matrix). This implies
that the
model is substantially different in CP violating sector from the standard
CKM model \cite{Buras:2003wd} where the KM phase must be in the
first quadrant. Such non-CKM CP violation, for instance, could be in the
squark mixings. Phenomenologically speaking, there is nothing wrong
with this possibility although admittedly it will require many random
parameters to reproduce the observed data on CP violation. 
%However, if recent indications of deviations from
%CKM CP
%violation in the $B\rightarrow \phi K_S$ are confirmed, one would
%certainly need to go beyond the CKM scenario. Also recall that there are
%independent reasons
%(such as solving strong CP problem, understanding the origin of matter
%etc) 
%to think that CP violating phenomena are more complex than perhaps CKM
%physics alone would indicate and one must seek additional experimental
%information to decide the nature of CP violation.

Since the CKM model is simple and seems
consistent with known data on CP violation and it would be interesting to
see if the above predictive model for neutrinos can coexist with the
simple CKM CP violation. One of the objectives of this paper is to study
this using the type I seesaw formula. It
was mentioned  in Ref.\cite{Goh:2003hf} that the main reason for the
constraint on CP phase being in the third quadrant in the pure type II model 
has to do with fitting
the electron mass and if a nonrenormalizable term is introduced to fix
this problem, then the CP phase could be of CKM type. 

In this paper, we focus primarily on the same minimal SO(10) model but 
with both type I and type II seesaw formula for neutrinos. We obtain the
following results: (i) We first show that
in a certain limit, one can get the relation $M_{\nu}\simeq c
(M_d-M_e)$,
so that the basic mechanism that led to the maximality of solar and
atmospheric neutrino mixings in the case of pure type II seesaw formula is
preserved in the type I case. As far as we know, this fact was not noticed
in any of the earlier papers that analyzed type I seesaw in minimal
SO(10). We then show that the model is in conflict with neutrino data
with or without CP violating phases. This is different from the pure type
II case where, one could get the neutrino predictions correctly within the
renormalizable framework without any CP phase in the Yukawa couplings
as well as the KM phase in second quadrant. (ii) Secondly, we show that we can 
restore compatibility with KM
CP violation for both type I and type II cases by using
the SO(10) model as an effective theory at the GUT scale, where we 
generate a specific nonrenormalizable term. This not only leads to
successful predictions for
neutrino mixings but also maintains the CKM model for CP violation.
Of course, the resulting model is not a minimal model anymore since at 
the Planck scale, it can emerge from a renormalizable theory with two
pairs of {\bf 126} fields (rather than one in the minimal case). 
(iii) We also find that the type I model
works only for values of tan$\beta$ higher than 30, providing a way to
test the model in supersymmetry experiments. (iv) We evaluate the
lepton flavor violating processes and 
electron electric dipole moment (EDM) by using the predicted
leptonic mass matrices in both type I and pure type II case and show that
they are in a range accessible to the present experiments. (v) We find
that for the type I model to work, 
%the B-L breaking 
intermediate scale ($v_R$) must be below
the GUT scale. We present a brief discussion of unification of gauge
couplings in the model to show that this can indeed happen. 

This paper is organized as follows:
in section 2, we give a basic framework
of the minimal SO(10) model and show that the relation
$M_{\nu}\, \simeq\, c(M_d-M_e)$ needed to understand large neutrino
mixings follows in a certain limit even with the type I or mixed type
II seesaw. We examine its predictions for the case where the 
 type I seesaw formula is used for discussing neutrino masses and
 KM phase is kept around 60-70 degrees as required in the standard
model fit. We find that the predictions for solar neutrino oscillation
do not agree with the current combined data analysis of mixing angles
and ratio of mass squared differences for this case. We also discuss the
restriction on $\tan\beta$ in  the mixed type II and type I models.
In section 3, to remedy the situation of neutrino masses and mixing angles,
we include specific types of non-renormalizable terms in the
superpotential and show how they
are helpful in reconciling the neutrino predictions with CKM CP violation for both
type I and type II models. We also discuss the origins of these terms.
In sec. 4, we discuss the predictions of the model giving numerical
results for various parameters as well as neutrino masses and mixings.
In section 5, we discuss the gauge coupling unification in the model and
show that the $B-L$ scale required for obtaining neutrino masses is
compatible with gauge coupling unification. In sec. 6, we evaluate the
leptonic flavor violation
and lepton electric dipole moment for both type I and type II models.
Section 7 contains our conclusions.

\section{Minimal Renormalizable SO(10) Model and Its Predictions}

In the minimal SO(10) model,
Yukawa interactions are given as the couplings of the $\mathbf{16}$-dimensional
matter spinor $\psi_i$ with only one $\mathbf{10}$-dimensional Higgs $H$
and one $\overline{\mathbf{126}}$ Higgs $\Delta$.
The superpotential for Yukawa interactions is written as
\begin{equation}
W_Y = \frac12 h_{ij} \psi_i \psi_j H + \frac12 f_{ij} \psi_i \psi_j
\Delta \ .
\label{SO(10) superpotential}
\end{equation}
The Yukawa couplings, $h$ and $f$, are complex symmetric $3\times 3$ matrices. 

The SO(10) gauge symmetry breaks down by the Higgs mechanism.
There are several breaking pattern in SO(10) GUT.
The SO(10) symmetry is broken to the left-right group, $G_{2231} =
SU(2)_L \times SU(2)_R \times SU(3)_c
\times U(1)_{B-L}$, using
the $\mathbf{54}$ and $\mathbf{210}$ Higgses.
To break $SU(2)_R \times U(1)_{B-L}$ down to $U(1)_Y$, we will use
%one can use either $\mathbf{16}+ \mathbf{\overline{16}}$ or 
$\mathbf{126}+\mathbf{\overline{126}}$ Higgs multiplets.
%In our model, the $\mathbf{126}+\mathbf{\overline{126}}$ Higgses
%are employed, and the $\mathbf{\overline{126}}$ is the $H_{126}$ which 
%couples to the fermions.
%The electroweak symmetry, $SU(2)_L \times U(1)_Y$, breaks down to $U(1)_Q$ by
%the $\mathbf{10}$ Higgs.

The SO(10)-invariant superpotential, Eq.(\ref{SO(10) superpotential}),
 includes MSSM Yukawa couplings plus
right handed Majorana mass term
which are written by using the MSSM superfields:
\begin{equation}
W_Y \supset Y_{ij}^u Q_i U^c_j H_u + Y_{ij}^d Q_i D_j^c H_d + Y_{ij}^e L_i E_j^c H_d
+ Y_{ij}^\nu L_i N_j^c H_u + \frac12 f_{ij} N^c_i N^c_j \Delta_R \ ,
\label{MSSM-superpotential}
\end{equation}
where $H_u$ and $H_d$ are MSSM Higgs doublets which are linear
combinations of the SM doublets in {\bf 10} and $\mathbf{\overline{126}}$ Higgs
multiplets, and $\Delta_R$ is  part of the $\mathbf{\overline{126}}$
 field $\Delta$. 
%and an $SU(2)_R$ triplet with $B-L$ charge.
We note that the VEV of $\Delta_R$ gives right handed neutrino masses
and breaks $SU(2)_R \times U(1)_{B-L}$ symmetry down to $U(1)_Y$.
%Standard Model singlet
%part of $\Delta$ whose vacuum expectation value (VEV) breaks $B-L$ symmetry.
As noted the SO(10) Higgs fields, $H$ and $\Delta$, contain two pairs of 
$SU(2)_L$ Higgs doublets, $H^{u,d}_{10}$ and $H^{u,d}_{126}$,
and the MSSM Higgs doublets are linear combinations of two pairs,
\begin{eqnarray}
H_u &=& \alpha_u H_{10}^u + \beta_u H_{126}^u, \\
H_d &=& \alpha_d H_{10}^d + \beta_d H_{126}^d,
\end{eqnarray}
where $|\alpha_{u,d}|^2 + |\beta_{u,d}|^2 =1$.
The other linear combinations are assumed to be massive around %$B-L$ breaking scale.
GUT scale.
%and there is one pair of MSSM Higgs doublets below the GUT scale.
The MSSM Yukawa couplings $Y^{u,d,e,\nu}$ are given by linear combination
of $h$ and $f$,
%\begin{eqnarray}
%Y^u &=& h \alpha_u^* + f \beta_u^* , \\
%Y^d &=& h \alpha_d^* + f \beta_d^* , \\
%Y^e &=& h \alpha_d^* -3 f \beta_d^*, \\
%Y^\nu &=& h \alpha_u^* -3 f \beta_u^*.
%\end{eqnarray}
and the fermion mass matrices and Majorana mass matrix for right handed
neutrino 
are given as\footnote{The mass matrices are defined as $-{\cal L}_m =
\overline{\psi_L} M \psi_R
+ \frac12 {(\nu_R)}^TC^{-1} M_R \nu_R$ + h.c.}
\begin{eqnarray}
M_u &=& (h^* \alpha_u + f^* \beta_u) v_u , \\
M_d &=& (h^* \alpha_d + f^* \beta_d) v_d , \\
M_e &=& (h^* \alpha_d -3 f^* \beta_d) v_d , \\
M_\nu^D &=& (h^* \alpha_u -3 f^* \beta_u) v_u , \\
M_R &=& f^* v_R,
\end{eqnarray}
where $v_{u,d}$ are VEVs of MSSM Higgs doublets
and $v_R$ is a VEV of $\Delta_R$.
We denote that $v_u = v \sin \beta$ and $v_d = v \cos \beta$,
where $v= 174$ GeV.
Then we have sumrules for leptonic mass matrices:
\begin{eqnarray}
M_e &=& c_d M_d + c_u M_u , \label{charged_lepton_mass}\\
M_\nu^D &=& \frac{c_d + 3}{c_u} (M_d- M_e) + M_u ,\label{dirac_neutrino_mass}\\
M_R &=& \frac{M_d - M_e}{4 \beta_d v_d} v_R ,\label{right_handed_neutrino_mass}
\end{eqnarray}
where 
\begin{equation}
c_u = \frac{4 \cot \beta}{\alpha_u/\alpha_d - \beta_u/\beta_d},\quad
c_d + 3 = \frac{-4 \beta_u/\beta_d}{\alpha_u/\alpha_d - \beta_u/\beta_d}.
\label{cd-and-cu}
\end{equation}
So if the quark mass matrices $M_u$ and $M_d$ are our input, then
leptonic mass matrices
are determined by two complex parameters, $c_u$ and $c_d$,
barring the overall scale and the phase of Majorana neutrino mass matrix,
$M_R$.
Thus, using the masses of  three charged leptons,
we predict the neutrino mass matrices in terms of only one real parameter. 
This is an interesting feature of the minimal SO(10) model and is
responsible for the high predictive power of the model.

Let us count the  number of parameters in the fermion mass matrices in
the minimal SO(10) model.
%in the sumrules in
%Eqs.(\ref{charged_lepton_mass}-\ref{right_handed_neutrino_mass}).
Rotating the fermion fields by a unitary matrix
without violating  SO(10) symmetry (fermions in $\mathbf{16}$ are all rotating
simultaneously),
up-type quark mass matrix, $M_u$, can be made real positive diagonal,
$M_u = $ diag$(m_u,m_c,m_t) \equiv D_u$.
Then down-type quark mass matrix, $M_d$, is a general complex symmetric matrix,
which has 6 complex parameters.
The matrix can be written as $M_d = U D_d U^T$ where $D_d$ is real positive diagonal,
$D_d \equiv $ diag$(m_d,m_s,m_b)$, and $U$ is a unitary matrix.
The unitary matrix has 9 real parameters (3 angles and 6 phases)
and we can parameterize the unitary matrix as $U= P_u^* \bar U P_d$.
The matrices $P_u$ and $P_d$ are diagonal phase matrices,
$P_u\equiv$ diag$(e^{i \phi_u/2},e^{i \phi_c/2},1)$ and
$P_d\equiv$ diag$(e^{i \phi_d/2},e^{i \phi_s/2},e^{i \phi_b/2})$
and $\bar U$ is just same as the Cabibbo-Kobayashi-Maskawa (CKM) matrix,
$V_{CKM}$,
in which there are 3 mixing angles and 1 phase (KM phase).
The phase matrices, $P_u$ and $P_d$, are unphysical in the MSSM quark sector
since they can be absorbed in right handed quark fields,
but the phases are relevant parameters in the leptonic mass matrices
through the sumrules in Eqs.(\ref{charged_lepton_mass}-\ref{right_handed_neutrino_mass}).
One overall phase of $M_d$ is irrelevant in the sumrules,
using the rotation, $M_d \rightarrow e^{i \vartheta} M_d$, $M_u \rightarrow M_u$.
This rotation is generated by $(\alpha_d,\beta_d) 
\rightarrow (e^{i \vartheta} \alpha_d, e^{i\vartheta} \beta_d)$.
So, we can choose the phase $\phi_b$ to be zero.
Finally, we have 19 parameters in the sumrules:
14 real parameters in quark mass matrices $M_u$ and $M_d$
(6 quark masses, 3 mixing angles, 1 KM phase,
and 4 unknown phases, $\phi_{u,c,d,s}$),
2 complex parameters, $c_u$ and $c_d$, and 1 overall scale parameter
in the Majorana mass matrix $M_R$ (overall phase of $M_R$ is not counted
since it is not physical).
On the other hand, the MSSM Yukawa couplings plus the right handed 
Majorana mass in Eq.(\ref{MSSM-superpotential}),
give us 32 parameters:
10 parameters in $Y^u$ and $Y^d$ (6 quark masses and 3 mixings and 1 KM phase),
10 parameters in $Y^e$ and $Y^\nu$ in the same way,
and then 12 parameters in the Majorana mass matrix which is 
generally a symmetric matrix.
So, in the minimal SO(10) model, we have less parameters than the MSSM.

The quark masses, mixings and the KM phase are our inputs.
We can redefine the basis of fermions, and the up- and down-type quark
mass matrices are written as
$M_u= D_u P_u^2$, $M_d= V_{CKM} D_d P_d^2 V_{CKM}^T$.
The authors in Ref.\cite{Matsuda:2000zp}
assumed that the phase matrices $P_{u,d}^2$ are real 
and they found a solution to generate the observed quark masses
and the CKM matrix and the charged lepton masses in the case
where $\phi_{d,s,c}=\pi$ and $\phi_u = 0,\pi$.
In our general analysis, we take those phases $\phi_{d,s,u,c}$ to be arbitrary.
Those phases are constrained to obtain the bi-maximal neutrino oscillation data.
%$\phi_u$ can be arbitrary,
%but the other phases $\phi_{d,s,c}$ are constrained 
%due to electron mass fitting.
%Furthermore, those phases are more constrained to obtain bi-maximal neutrino
%oscillation \cite{Goh:2003hf}.
%The elements of the phase matrices $P_u$ and $P_d$ are free parameters,
%but they are constrained by fitting neutrino oscillation data.

Let us see how one can obtain a numerical fit of the parameters in
the lepton sector at the GUT scale.
The charged lepton mass matrix is written as
$M_e= c_d M_d + c_u M_u$.
So, 
$\tau$ and $\mu$ masses are approximately (neglecting generation mixing and phases),
\begin{equation}
\pm m_\tau \simeq c_d m_b + c_u m_t, \qquad \pm m_\mu \simeq c_d m_s + c_u m_c.
\label{appro_tau}
\end{equation}
Using the relations $m_c/m_t \ll m_s/m_b$ and $m_\tau \simeq m_b$,
we obtain
$c_d \simeq \pm m_\mu/m_s$ and $c_u \simeq - m_b/m_t (c_d \pm 1)$,
and thus $|c_d|$ is about 3, naively.
% the muon mass is approximately $m_\mu \simeq |c_d| m_s$
%and $|c_d|$ is about 3, naively.
In this naive approximation, however, the electron mass becomes about $c_d m_d$,
which is too large,
so we cannot neglect the off-diagonal elements,
and we need a fine-tuning in det$(M_d + \kappa M_u)$,
where $\kappa \equiv c_u/c_d$,
by a suitable choice of $\kappa$.
The numerical  fit is given in Ref.\cite{Matsuda:2000zp}.
The key ingredient to fit the electron mass is the 
strange quark mass.
Actually, the current strange quark mass still has large experimental error and
we can make it to be a parameter in the model.
We show the values of the strange quark mass (at 1 GeV) and $|c_d|$ by
varying the KM phase in Fig.1. 
%The usual KM phase $\delta$ is defined in the range, $0\leq \delta \leq 2\pi$.
We take the KM phase to be in the first and second quadrant without loss of generality,
since the sumrule does not change under the conjugation.
%When the KM phase is 60-70 degree, which is favored in experiment,
%the strange quark mass is minimal and $|c_d|$ is about 6-7.
The three real parameters are consumed to fix the three charged lepton masses:
$\kappa$ and $|c_d|$ are determined and the phase of $c_d$ is still
undetermined.

\begin{figure}[tbp]
\centering
\includegraphics*[angle=0,width=6.2cm]{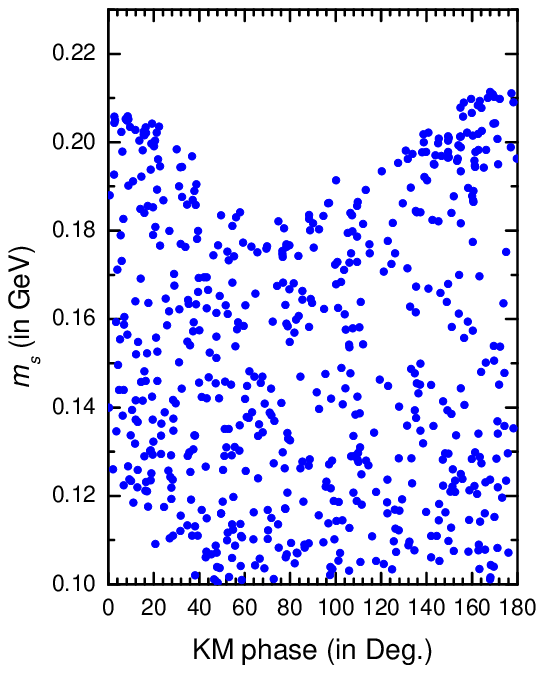}
\includegraphics*[angle=0,width=6cm]{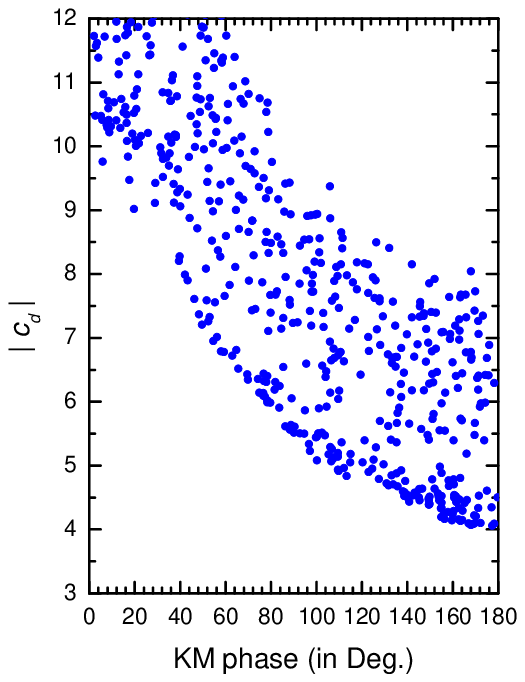}
\caption{The strange quark mass (at 1 GeV) and $|c_d|$ to fit charged lepton masses are shown.}
\label{Fig.1}
\end{figure}

We now determine the neutrino mass matrix.
First, in the pure-type II case,
the light neutrino mass, ${\cal M}_\nu^{\rm II}$, is given as $f^* v_L$ which 
is proportional to $M_d - M_e$.
The charged-lepton mass matrix is written as $M_e= c_d (M_d + \kappa M_u)$.
%Using Wolfenstein parameterization of CKM matrix,
We define the complex number $\xi$ as
\begin{equation}
\xi \equiv |c_d| (M_d + \kappa M_u)_{33}/m_\tau.
\end{equation}
By definition, $|\xi| \simeq 1$. 
The phase of $\xi$ is determined by the charged lepton mass fitting.
The (3,3) element of $M_e$ is then $\xi e^{i\sigma} m_\tau$,
where $\sigma$ is a phase of $c_d$, $e^{i\sigma}= c_d/|c_d|$.
We denote $\hat \xi \equiv \xi e^{i \sigma}$.
The pure type II neutrino mass matrix is given as 
\begin{eqnarray}
{\cal M}_\nu^{\rm II} &\propto& (1-c_d) M_d - c_u M_u \nonumber \\
&\simeq& (1-c_d)
      \left(
        \begin{array}{ccc}
                m_d e^{i\phi_d} + V_{us}^2 m_s e^{i\phi_s} &
                   V_{us} m_s e^{i\phi_s} & V_{ub} m_b \\
                V_{us} m_s e^{i\phi_s} & m_s e^{i \phi_s} & V_{cb} m_b \\
                V_{ub} m_b & V_{cb} m_b  & (m_b - \hat \xi m_\tau)/(1-c_d)
               \end{array}
      \right).
\end{eqnarray}
In the expression, $c_u m_c$ and $c_u m_u$ terms in the diagonal element
are neglected.
Choosing the phase $\sigma$ to make $\hat \xi \simeq 1$,
the (3,3) entry of the matrix is $m_b-m_\tau$.
Since $b$-$\tau$ mass convergence implies that $m_b$ and $m_{\tau}$ become
close to each other as we move to the GUT scale, we can assume that 
$m_b - m_\tau \sim (1- c_d) V_{cb} m_b$, which implies that
 the neutrino mass matrix is given as \begin{equation}
{\cal M}_\nu  \sim \left(
               \begin{array}{ccc}
                \lambda^2 & \lambda & \lambda \\
                \lambda & 1 & 1 \\
                \lambda & 1 & 1
               \end{array}
         \right) m_0 ,
\label{bi-large}
\end{equation}
where $\lambda \simeq V_{us} \sim 0.2$.
We then obtain bi-large neutrino mixing: both atmospheric and
solar angle are large.
The third neutrino mass, $m_{\nu_3}$, is about $m_0$.
To obtain the large solar mixing,
determinant of (2-3) block of the neutrino matrix is required to be
less than $\lambda m_0^2$,
and then the second neutrino mass, $m_{\nu_2}$, is about $\lambda m_0$.
So the mass squared ratio is predicted as
$\Delta m^2_{\odot}/\Delta m^2_A \sim \lambda^2$.
The 1-3 neutrino mixing is also predicted as $U_{e3} \sim \lambda$.
%It is interesting that the bottom-tau unification (but not exact unification)
%gives large atmospheric neutrino mixing.

Next we consider the case of mixed type II neutrino mass matrix.
This case is more complicated than the pure type II case. From
Eqs.(\ref{dirac_neutrino_mass}--\ref{right_handed_neutrino_mass}),
the mixed type II seesaw matrix is given as
\begin{equation}
{\cal M}_\nu^{\rm{II\, mixed}} \propto
(M_d - M_e)(1+\Delta) + 2 \frac{c_u}{c_d+3} M_u + \left(\frac{c_u}{c_d+3}\right)^2 M_u (M_d-M_e)^{-1} M_u.
\label{mixed-type-II}
\end{equation}
The first term is just the same as in  the pure type II case. When $\Delta=0$, 
we get the 
type I case. 
The second and third terms might give a hierarchical structure to the (2-3) block
of neutrino mass matrix, and would spoil large atmospheric mixing.
The (3,3) element of the Eq.(\ref{mixed-type-II}) can be written approximately as
\begin{equation}
\left(
(1-\hat \xi) (1+\Delta)
+ 2 \frac{\hat \xi - c_d}{c_d + 3} + \zeta \left(\frac{\hat \xi - c_d}{c_d + 3} \right)^2
\right) m_\tau,
\label{mixed-33}
\end{equation}
where $\zeta = [(M_d-M_e)^{-1}]_{33} m_\tau$.
This element is of the order of $m_\tau$ or larger in general.
However, if the (3,3) element
%\begin{equation}
%\frac{c_u}{c_d+3} m_t \left(
% \frac{c_u}{c_d+3} m_t (M_d-M_e)^{-1}_{33} + 2 \right),
%\label{3-3 element}
%\end{equation}
can be canceled to  the order of  $(1+\Delta)(1-c_d) V_{cb} m_\tau$,
then the bi-large mixing neutrino matrix can be realized.
Assuming that $\hat \xi \simeq 1$  as in pure type II case,
the cancellation condition of the second and third terms is
$c_d = -(6+\zeta)/(2-\zeta)$.
In our numerical analysis,
the parameter $\zeta$ is almost real and $\zeta \simeq 1$.
Thus, the cancellation condition is $c_d \sim -7$.
At that time, the phase of $c_d$ is almost $\pi$,
and then the phase of $\xi$ must be almost $\pi$ from the condition $\hat \xi \simeq 1$.
The condition can be satisfied 
if the strange quark has a smaller value of mass,
since $|c_d| \sim m_\mu/m_s$ naively.
According to our numerical analysis,
$\hat \xi$ is not necessary to be 1
and $c_d$ can be different values from $-7$.
However, $|c_d|$ has a lower bound, $|c_d|\agt 5$ to cancel 
the (3,3) element of the neutrino mass matrix in our analysis.
Consequently, smaller value of strange mass is favored in the mixed type II case.

%By the numerical studies,
%the condition of the cancellation needs $c_d \sim -(5-7)$.
%Actually, we have an approximate relation, $c_u m_t \sim (|c_d|-1) m_\tau$
%and $m_\tau (M_d-M_e)_{33}^{-1} \sim O(1)$.
%We see that from the Fig.1, the cancellation can occur in the case 
%where the KM phase is about 50-90 degree,
%and the strange quark mass is at the lowest possible value in the figure.
%If the KM phase becomes large and is in the second quadrant,
%$|c_d|$ decreases
%and then the cancellation of (3,3) element does not occur.
%%This result is good agreement with observed KM phase and strange mass.
%The detailed numerical studies of type I seesaw given in Ref.\cite{Fukuyama:2002ch}
%agree with the qualitative picture above.

\begin{figure}[tbp]
\centering
\includegraphics*[angle=0,width=8cm]{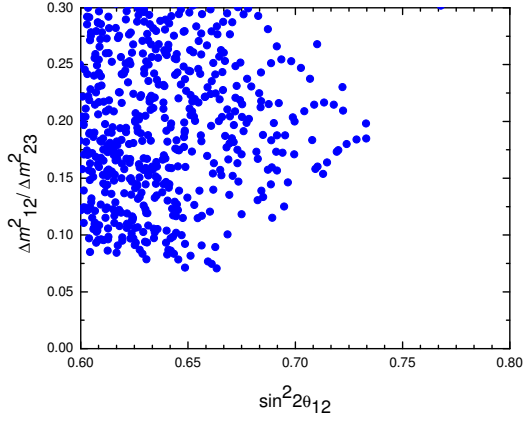}
\caption{The solar mixing angle and ratio of mass squared differences 
are dot-plotted for 
$\sin^2 2\theta_{23} > 0.89$. The allowed region from recent data fitting lies 
in the right-bottom corner in the figure.}
\label{Fig.2}
\end{figure}

In both pure and mixed type II (and of course in the type I case),
the neutrino mass matrix structure in Eq.(\ref{bi-large})
which gives bi-large mixing can be obtained.
%\begin{equation}
%{\cal M}_\nu \sim \left(
%               \begin{array}{ccc}
%                \lambda^2 & \lambda & \lambda \\
%                \lambda & 1 & 1 \\
%                \lambda & 1 & 1
%               \end{array}
%         \right) m_0 \ . \label{bi-large}
%\end{equation}
Then we have a good prediction of the 1-3 mixing angle, $|U_{e3}| \simeq \lambda$.
%The rough estimate of the 1-3 mixing is that $|U_{e3}| \sim V_{us} m_s/(V_{cb} m_b)$.
Actually, in the numerical studies in Ref.\cite{Goh:2003hf} (type II) 
and Ref.\cite{Fukuyama:2002ch} (type I),
the prediction of $|U_{e3}|$ is about 0.17-0.18.
Another important prediction of the minimal SO(10) model
is the solar mixing angle and the ratio of mass squared differences,
$\Delta m^2_{\odot}/\Delta m^2_A$.
%The mass squared ratio is determined by the eigenvalues of the neutrino mass matrix
The predictions are almost determined by the (2-3) block of the matrix.
The (2-3) block is approximately written as
\begin{equation}
{\cal M}_\nu^{{(2\mbox{-}3)}} \propto \left(
                          \begin{array}{cc}
                            m_s e^{i\phi_s} & V_{cb} m_b \\
                            V_{cb} m_b & \epsilon m_\tau
                          \end{array}
                             \right),
\end{equation}
where $\epsilon$ is a cancellation factor, 
$\epsilon m_\tau \simeq (M_d-M_e)_{33}/(1-c_d)$
in the pure type II case,
and $\epsilon m_\tau$ is approximately
Eq.(\ref{mixed-33}) divided by $1-c_d$ in the mixed type II case.
The $|\epsilon|$ should be less than $V_{cb} \sim \lambda^2$ 
to give rise to a large atmospheric mixing.
The condition to obtain a large maximal mixing angle is 
that the determinant of the (2-3) block is canceled,
%The condition is 
\begin{equation}
|\epsilon e^{i \phi_s} m_s m_\tau - V_{cb}^2 m_b^2| \alt O(\lambda)V_{cb}^2 m_b^2  .
\end{equation}
This condition also provides a small mass differences squared ratio,
$\Delta m^2_{\odot}/\Delta m^2_A \sim \lambda^2$.
%This is the interesting fact of the bi-large mixing structure.
To satisfy the condition, one needs $\epsilon e^{i\phi_s} \simeq + |\epsilon|$,
and $|\epsilon|$ can not be very small.
Then, maximal atmospheric mixing requires a relation
%\begin{equation}
$m_s - |\epsilon| m_\tau \ll 4V_{cb} m_b$.
%\end{equation}
Assuming $m_s \simeq |\epsilon| m_\tau$,
we obtain the condition for the bi-maximal neutrino mixing and the small mass
squared ratio as
\begin{equation}
m_s \simeq V_{cb} m_b.
\end{equation}
{}From the experimental data, 
the strange/bottom mass ratio at the GUT scale is always smaller
than $V_{cb}$,
thus a larger strange mass gives rise to larger mixing angles in the
minimal SO(10) model.
Furthermore, in the minimal model,
the strange quark mass is constrained due to electron mass fitting.
As we can see in Fig.1,
the maximal values of the strange mass depend on the KM phase,
and the second quadrant KM phase gives larger
strange quark mass.
As a result, 
the bi-maximal neutrino mixings and the small mass squared ratio
favor the second quadrant KM phase.
This result is in agreement with the numerical studies in
Ref.\cite{Goh:2003hf}.
In the case of mixed type II or type I,
smaller values of strange quark mass are favored to
obtain the cancellation of the (3,3) component of the neutrino mass matrix
as we have noted.
So in  type I and mixed type II case, there is a tension between
getting large neutrino mixings in general and getting their precise values
as well as the desired mass square ratio. As a result, the minimal
SO(10) model with type I or mixed type II predicts
that the solar mixing angle is not very large and the mass squared ratio
is not very small as long as the atmospheric neutrino mixing is maximal.
We give our numerical data plotting for the solar mixing angle and mass squared
ratio in the Fig.\ref{Fig.2}.
In the mixed type II or type I case,
the mixing angle and mass squared ratio are bounded for any KM phase, and 
%In our numerical studies, we obtain the bounds for the solar mixing angle
%and mass squared ratio in the type I case
%as in the Fig.\ref{Fig.2}
%:  %where the KM phase is in the first quadrant:
%\begin{equation}
%\sin^2 2 \theta_{12} < 0.75, \qquad \Delta m^2_\odot/\Delta m^2_A > 0.1,
%\end{equation}
%
%as long as $\sin^2 2\theta_{23} > 0.89$.
%The allowed region from recent data fitting lies in the right-bottom corner
%in the figure,
recent data fitting excludes such bounds more than $2\sigma$ level
\cite{Barger:2003qi}.

%This fact is supported by the numerical studies in Ref.\cite{Fukuyama:2002ch}.
%In the case of pure type II,
%we can make the KM phase to be in second quadrant
%and then solar mixing angle can be large and mass squared ratio can be small.
%Such result is shown by numerical studies in Ref.\cite{Goh:2003hf}.

Next we will see the $\tan\beta$ bound in the minimal SO(10) model. From 
Eq.(\ref{cd-and-cu}),
we obtain
\begin{equation}
\frac{\alpha_u}{\alpha_d} = \frac{1-c_d}{c_u} \cot \beta, \qquad 
\frac{\beta_u}{\beta_d} = -\frac{c_d+3}{c_u} \cot \beta.
\end{equation}
Since $\alpha_{u,d}$ and $\beta_{u,d}$ are Higgs mixings,
we have unitarity constraint $|\alpha_{u,d}|^2 + |\beta_{u,d}|^2 =1$.
Thus we have $(|\alpha_u/\alpha_d|-1) (|\beta_u/\beta_d|-1)\leq 0$,
and 
\begin{equation}
{\rm min} \left(\left|\frac{c_d+3}{c_u}\right|,
\left|\frac{1-c_d}{c_u}\right|
\right) \leq \tan\beta
\leq 
{\rm max} \left(\left|\frac{c_d+3}{c_u}\right|,
\left|\frac{1-c_d}{c_u}\right|
\right).
\end{equation}
In the case where $\hat \xi \simeq 1$ which is favored to obtain bi-large structure,
$(1-c_d)/c_u \simeq m_t/m_b$,
and then we have lower bounds of $\tan\beta$,
such as 
$\tan\beta \agt m_t/m_b |(c_d+3)/(1-c_d)|$.
In the case where $c_d \sim -7$ which is favored in the mixed type II (and type I) case,
the lower bound of $\tan\beta$ is roughly 30.
To obtain small $\tan\beta$ such as $\tan\beta \sim 10$,
$c_d \sim -3.5$ is needed.
Such small $|c_d|$ can happen only in the case
where the  KM phase is close to 180 degree in the minimal SO(10) model.

In the type I seesaw case,
the scale of the right handed Majorana mass is  predictable.
Using the Eqs.(\ref{dirac_neutrino_mass}-\ref{right_handed_neutrino_mass}),
the light neutrino mass scale, $m_0$ in Eq.(\ref{bi-large}),
can be written as
\begin{equation}
m_0 \simeq (1- c_d) \left(\frac{c_d+3}{c_u}\right)^2 V_{cb} m_b \frac{4 \beta_d v_d}{v_R}.
\end{equation}
So using  the mass squared magnitude from the  atmospheric neutrino data,
we can fix the mass parameter, $m_0$,
and obtain the right handed Majorana mass matrix,
Eq.(\ref{right_handed_neutrino_mass}).
The lightest Majorana mass, which is an important scale for leptogenesis,
is about $10^{11}$ GeV.
Also we obtain the upper bound of $v_R$ since $|\beta_d|$ is less than 1. 

Now let us summarize the predictions of the minimal SO(10) model.
In both pure type II and mixed type II,
we can obtain the bi-large mixing structure, Eq.(\ref{bi-large}),
for the solar and the atmospheric oscillation.
In the bi-large neutrino matrix, the 1-3 mixing is well predicted as
$|U_{e3}|\simeq 0.17-0.18$.
In the mixed type II (and type I) case,
larger $|c_d|$ is favored to obtain bi-large structure in the neutrino mass
matrix.
In order to obtain both bi-maximal neutrino mixing and small mass squared ratio,
a larger strange quark mass is favored,
and this requires a smaller  $|c_d|$. 
So, in the mixed type II (and type I) case,
the bi-maximal neutrino and the small mass squared ratio
are not favored,
and it is hard to achieve a best fit value of the recent combined data
analysis. In the pure type II case, 
the condition of bi-large mixing is just $\hat \xi \simeq 1$,
and $|c_d|$ is not constrained.
However, due to the electron mass fitting, 
the strange mass is related to the KM phase,
and if the KM phase is 60-80 degree,
the strange mass can not be large
and thus it is hard to achieve the  
bi-maximal mixing and the small mass squared ratio in the pure type II case as well.
If the KM phase is in the second quadrant 
and especially for the larger value of KM phase,
the strange quark mass can be large and
the best fit values for neutrino oscillation can be achieved.
The $\tan\beta$ in the mixed type II and type I model is greater than
$30$ roughly and the ${B-L}$
breaking scale has an upper-bound in the type I model.

\section{Inclusion of Nonrenormalizable terms}

%In order to reproduce the CKM model of CP violation at low energies,
%we employ corrections to the SO(10) model which originate from
%non-renormalizable terms.
%The most simple non-renormalizable term is written including an extra
%singlet $S$
%\begin{equation}
%W_Y^{\rm nr} = 
%\frac12 \frac{S}{M_P} ( h_{ij}^\prime \psi_i \psi_j H_{10} +  f_{ij}^\prime 
%\psi_i \psi_j H_{126}) \ .
%\end{equation}
%In this case, there is no change from the results of the renormalizable
%case since, the effect of these terms is simply to redefine the couplings
%$h$ and $f$ of Eq. (3). We therefore seek other renormalizable terms.

As we have seen in the previous section,
the fitting of three charged-lepton masses (especially, electron mass)
gives relatively small strange quark mass in the case where
the KM phase is 60-80 degree.
Then the (2,2) element of the neutrino mass matrix is relatively smaller
than (2,3) element,
and as a result, we get bounds for the solar mixing angle and 
the mass squared ratio. 
%
%To obtain larger solar mixing, we need to increase the strange
%quark mass in the (2,2) element
%of $M_e$ while keeping $|\det M_e| = m_e m_\mu m_\tau$.
%If we switch on the (1,3) element and/or (1,1) element
%of $\Delta M_e$ matrix,
%we can increase the strange quark mass.
%(Precisely speaking, what we need is to increase
% $(M_e)_{22}/c_d$.
%If we switch on the (2,2) element of $\Delta M_e$,
%the ``effective" strange mass is modified
%to $m_s + (\Delta M_e)_{22}/c_d$.)
%%In the case of pure type II,
%%we can easily obtain the large solar mixing angle and the  
%%small mass squared ratio
%%with the KM phase in the first quadrant.
%The point is that when the strange quark mass increases, $|c_d|$ decreases
%since $c_d m_s$ is almost fixed by the muon mass input. 
%Then the cancellation of the (3,3) element of type I neutrino mass,
%Eq.(\ref{3-3 element}),
%becomes non-trivial.
%To obtain a bi-maximal mixing in type I seesaw,
%we need to keep $|c_d|$ to be 6-7
%and to increase the effective strange mass in the (2,2) element of
%neutrino mass matrix.
%
In this section, 
%we propose a way to remedy the
%situation by including a simple
%non-renormalizable term in the superpotential, that can
%make the (2,2) element larger. 
in order to reproduce the CKM model of CP violation at low energies,
we employ corrections to the SO(10) model which originate from
non-renormalizable terms in the superpotential.

The most simple non-renormalizable term is written including an extra
singlet $S$
\begin{equation}
W_Y^{\rm nr} = 
\frac12 \frac{S}{M_P} ( h_{ij}^\prime \psi_i \psi_j H_{10} +  f_{ij}^\prime 
\psi_i \psi_j H_{126}) \ .
\end{equation}
In this case, there is no change from the results of the renormalizable
case since the effect of these terms is simply to redefine the couplings
$h$ and $f$ of Eq.(3). We therefore seek other renormalizable terms.

We will employ {\bf 210} Higgs multiplet
to remedy the situation. 
Note that {\bf 210} Higgs (to be denoted $\Sigma$) can
be used to break the
SO(10) symmetry down to $SU(3)_c\times SU(2)_L\times SU(2)_R\times
U(1)_{B-L}$ if we give the $\mathbf{(1,1,1)}$ and $\mathbf{(1,1,15)}$ (under $G_{224}$)
submultiplets of {\bf 210} vev. One can then include the term
\begin{equation}
\frac12 \frac1{M_P} h^{\prime}_{ij} \psi_i \psi_j H \Sigma
\end{equation}
Since $\Sigma$ has four SO(10) indices, the combination $H\Sigma$ that can
couple to spinors is either a {\bf 120} or {\bf 126} representation. We
restrict the high scale theory (ultraviolet completion) from which the
nonrenormalizable term originates in such a way that no {\bf 120} term in
the effective $H\Sigma$ term couples to matter. For instance, we could
add a second {\bf 126} + $\overline{\bf 126}$ pair of field with a high
mass $M\gg M_U$ which does not develop a VEV with a superpotential given 
as follows\footnote{For another nonminimal version with multiple {\bf
126} at the GUT scale, see Ref.\cite{chen}.}:
\begin{eqnarray}
W'~=~h' \psi \psi\bar{\Delta}' + M\bar{\Delta}'{\Delta}'+ \Delta' \Sigma H
\end{eqnarray}
The effective theory below the scale $M$ has a nonrenormalizable term of the
type in Eq. (30) where only {\bf 126} part of $\Sigma H$ field product
contributes.
If only  $\mathbf{(1,1,1)}$ and $\mathbf{(1,1,15)}$ submultiplets of {\bf
210} acquire vev, the effective operator that contributes to
fermion masses then has $\mathbf{(2,2,15)}$ quantum numbers under
$G_{224}$. This
leads to mass formulae for the quarks and leptons of the following type.
\begin{eqnarray}
M_u &=& (h^* \alpha_u + f^* \beta_u) v_u + h^{\prime *}\alpha_u v_u \langle \Sigma \rangle/M_P, \\
M_d &=& (h^* \alpha_d + f^* \beta_d) v_d +h^{\prime *}\alpha_d v_d\langle \Sigma \rangle/M_P, \\
M_e &=& (h^* \alpha_d -3 f^* \beta_d) v_d -3h^{\prime *}\alpha_d v_d\langle \Sigma \rangle/M_P, \\
M_\nu^D &=& (h^* \alpha_u -3 f^* \beta_u) v_u -3h^{\prime *}\alpha_u v_u\langle \Sigma \rangle/M_P, \\
M_R &=& f^* v_R,
\end{eqnarray}
where $v_{u,d}$ are VEVs of MSSM Higgs doublets
and $v_R$ is VEV of $\Delta_R$. Note also that $h'$ is not a symmetric matrix
in general, but we assume that $h'$ is a symmetric matrix like $h$ and $f$ for simplicity.

We can now rewrite the above equations as follows 
%by denoting $-3h'\alpha_d v_d = \Delta M_e$.
\begin{eqnarray}
M_e &=& M_e^0 + \Delta M_e,\\
M_\nu^D &=& \frac{c_d + 3}{c_u} (M_d- M_e^0) + M_u + \frac{1-c_d}{c_u} \Delta M_e ,\\
M_R &=& \frac{M_d - M_e^0}{4 \beta_d v_d} v_R ,
\end{eqnarray}
where $M_e^0 = c_d M_d + c_u M_u$ and 
$\Delta M_e = -4 h^{\prime *} \alpha_d v_d \langle \Sigma \rangle/M_P$.
In this sumrules,
we can break the relation of the (2,2) element between charged-lepton and neutrino mass
matrices,
and thus we can increase (2,2) element of seesaw neutrino mass matrix.
%In pure type II, we can obtain solutions if we include $\Delta M_e$.
In the mixed type II (and type I) case, we have to increase the (2,2) element
while keeping $|c_d|$ to be larger than 5.
In fact, we find a large solar mixing solution in type I
when we switch on the (2,2) element of the $\Delta M_e$.

In the pure type II case, we do not need the (2,2) element of $\Delta M_e$
since a small $|c_d|$ can give rise to maximal mixings. Instead, we need the
(1,1) element to cancel and to give the desired electron mass and
we then find a large solar mixing solution in the case where KM phase is 
in the first quadrant.

\section{Numerical Results}

As already noted, the minimal SO(10) model is predictive because
fermions have only two Yukawa couplings, one with  $\mathbf{10}$
and one $\overline{\mathbf{126}}$ Higgses. This leads to
 a sumrule for the leptonic mass matrices which determine the 
the Dirac neutrino and right handed Majorana mass matrices from the 
observed neutrino data.
In a generic model for neutrino masses, the neutrino  oscillation data
gives us information about the 
neutrino masses and mixing angles of light neutrino mass matrix,
but no information about the Dirac neutrino and the right handed Majorana mass
matrices separately. On the other hand,
 a separate information of the Dirac neutrino and the right handed Majorana mass
matrices are important to extract the predictions of the models for
 lepton flavor violation and leptonic CP violation.
In the minimal SO(10) model however,
these matrices are completely determined.

In  type I case that we are considering in this paper, we get:
\begin{equation}
Y_{\nu}=\left(\begin{array}{ccc} -0.00159-0.00014i & 0.00067-0.0036i &
 0.017 +0.015i \\ 0.00369-0.00026i & 0.0182+0.0107i &
-0.046-0.0228i \\ -0.022+0.0085i & -0.02-0.0477i & 0.58+0.126i\end{array}\right),
\end{equation} and 
\begin{equation}
M_R=\left(\begin{array}{ccc} 0.00059+0.000048i & -0.00022+0.0012i &
 -0.0058-0.0051i \\-0.00022+0.0012i& -0.0014-0.0063i &
0.011-0.013i \\ -0.0058-0.0051i&0.011-0.013i &
-0.037-0.0086i\end{array}\right)\times10^{14.08}\ {\rm GeV}.
\end{equation} 
The above matrices are calculated for $\tan\beta=40$ and are in  the basis
where the charged lepton masses 
are diagonal. The scale $10^{14.08}$ GeV is 
%the scale where $G_{2231}$ is broken down to the SM 
the VEV of $\overline{\mathbf{126}}$ Higgs which couples to fermions
and this magnitude of the scale is the maximum 
 possible value given  the inputs of the quark masses and the CKM mixings. The
$h$ and $f$ in the basis where $M_u$ diagonal are given as follows:
 \begin{equation}
h^*=\left(\begin{array}{ccc} 0.0012-0.0002i & 0.00067-0.00014i &
 0.0011  +0.0035i \\ 0.00067-0.00014i & -0.0029-0.00122i &
-0.037+0.0043i \\ 0.0011+0.0035i & -0.037+0.0043i & 1.15+0.129i\end{array}\right),
\end{equation}
\begin{equation}
f^*=\left(\begin{array}{ccc} -0.0012-0.00006i & -0.0025+0.000062i &
 0.00088 -0.0021i \\ -0.0025+0.000062i & -0.0097+0.00061i &
0.021-0.0011i \\0.00088 -0.0021i&0.021-0.0011i&
-0.0012-0.0345i\end{array}\right).
\end{equation}
The neutrino mixing angles and mass squared ratio are given as:
\begin{equation}
\sin^2\theta_{12}=0.87,\,\,\sin^2\theta_{23}=0.92,\,\,|U_{e3}|=0.22,\,\,\Delta
m^2_\odot/\Delta m^2_A=0.051.
\end{equation}

At $\tan\beta=50$, the same matrices are given by:
\begin{equation}
Y_{\nu}=\left(\begin{array}{ccc} -0.0017 - 0.0004i& 
    0.0012 - 0.0039i& 
    0.016+ 
      0.019i\\0.004 - 
      0.00056i& 
    0.0194 + 0.011i & -0.053 - 
      0.023i\\ -0.023 + 
      0.01i& -0.025 - 0.052i& 
    0.660 + 0.13i\end{array}\right),
\end{equation} and 
\begin{equation}
M_R=\left(\begin{array}{ccc} 0.00117 + 0.00026i& -0.00076 + 
      0.0024i& -0.010- 
      0.012i\\-0.00076 + 
      0.0024i& -0.00099 - 0.0118i& 
    0.0275 - 0.0234i\\-0.010 - 
      0.012i& 
    0.0275- 0.0234i& -0.066- 
      0.025i
\end{array}\right)\times10^{13.82}\ {\rm GeV}.
\end{equation} 
The neutrino mixing angles and mass squared ratio are given as:
\begin{equation}
\sin^2\theta_{12}=0.81,\,\,\sin^2\theta_{23}=0.90,\,\,|U_{e3}|=0.22,\,\,\Delta
m^2_\odot/\Delta m^2_A=0.06.
\end{equation}  

In the  pure type II case, we get :
\begin{equation}
Y_{\nu}=\left(\begin{array}{ccc} 0.0008-0.001i & -0.001-0.0057i &
 0.004-0.026i\\0.003+0.0013i& -0.0056+0.00015i&
-0.053-0.008i \\ -0.000039+0.0014i & -0.012+0.00096i& 0.627+0.1019i\end{array}\right),
\end{equation} and 
\begin{equation}
M_R=\left(\begin{array}{ccc} -0.0016-0.000047i & -0.0036-0.000029i &
 0.00077+0.0028i\\-0.0036-0.000029i& -0.014-0.00044i&
0.026+0.0008i \\ 0.00077+0.0028i & 0.026+0.0008i& -0.0158-0.0078i
\end{array}\right)\times10^{16.24}\ {\rm GeV}.
\end{equation} 
The above matrices are calculated for $\tan\beta=50$ and are in the basis
where the charged lepton masses
are diagonal. The scale $10^{16.24}$ GeV is the scale where SO(10) gets
broken to
$G_{2231}$ which subsequently breaks down to the SM. 
The neutrino mixing angles and mass squared ratio are given as:
\begin{equation}
\sin^2\theta_{12}=0.85,\,\,\sin^2\theta_{23}=0.91,\,\,|U_{e3}|=0.22,\,\,\Delta
m^2_\odot/\Delta m^2_A=0.084.
\end{equation}

At  $\tan\beta=40$, the same matrices are:
\begin{equation}
Y_{\nu}=\left(\begin{array}{ccc} -0.00027 - 0.00066 i & -0.0046 + 0.00089i &
 -0.023 - 0.0023i \\ -0.00073 + 0.0028i& 0.00074 - 
  0.0037i& 0.007- 
  0.052i  \\ -0.0008 - 0.0001i& 0.0013 - 
  0.0079i& -0.103 + 0.57i \end{array}\right),
\end{equation} and 
\begin{equation}
M_R=\left(\begin{array}{ccc} -0.00078+1.5\cdot 10^{-6}i & -0.0018+0.00040i &
 0.00039+0.0012i \\-0.0018 + 0.000040i& -0.007-8.7\cdot 10^{-6}i &
0.012+  7.8\cdot10^{-6}i\\  0.00039+0.0012i&0.012+ 
    7.8\cdot10^{-6}i&
-0.0081 - 0.00013i\end{array}\right)\times 10^{16.24}\ {\rm GeV}.
\end{equation} 
The neutrino mixing angles and mass squared ratio are given as:
\begin{equation}
\sin^2\theta_{12}=0.85,\,\,\sin^2\theta_{23}=0.92,\,\,|U_{e3}|=0.20,\,\,\Delta
m^2_\odot/\Delta m^2_A=0.058.
\end{equation}

Using these fits, we will calculate the lepton  flavor violating
processes and
the amount of leptonic CP violation in  type I and pure type II
scenarios in the mSUGRA model in sec. 6. 

Even though all the examples we have given have large $U_{e3}$, its value
 can be much smaller due to the presence of new parameters in the higher
dimensional term.

\section{Gauge Coupling Unification}
{}From the results of the previous section (Eq. (40) and (45)), we see that
for the type I model to be successful in predicting neutrino masses, the
intermediate scale ($v_R$) must be below the GUT scale and around $10^{14}$ GeV for
our examples. We assume the
resulting symmetry breaking chain is of type, $SO(10)\rightarrow
SU(3)_c\times SU(2)_L\times SU(2)_R \times U(1)_{B-L} \rightarrow
SU(3)_c\times SU(2)_L\times U(1)_Y$. This is very different from the gauge
coupling unification scenario in MSSM. It is therefore important check if
the type I models are compatible with gauge coupling unification. 

\begin{figure}[tbp]
    \centering
    \includegraphics*[angle=0,width=6cm]{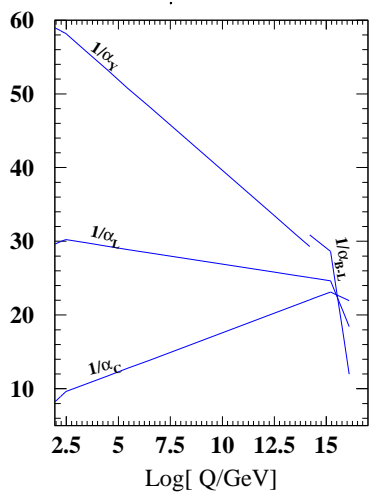}
    \vspace{0.1cm}
    \caption{\label{fig:fig0}  The couplings $1/\alpha_i$ are plotted as
a function of scale.}
\label{Fig.3}
\end{figure}

In our model, in the scale region $M_{SUSY} \leq \mu \leq v_{R}$, the
spectrum is that of familiar MSSM. Above the $v_{R}$ scale, the symmetry
group expands to $SU(3)_c\times SU(2)_L\times SU(2)_R \times U(1)_{B-L}$.
In addition to the new gauge bosons associated with this new symmetry, the
new matter and Higgs that contribute are the following: three RH
neutrinos as part of the $SU(2)_R$ lepton doublet; one bidoublet from
which the MSSM doublets emerged. In addition we need either the $B-L = 2$ triplet
pair ($\Delta_L +
\bar\Delta_L \oplus \Delta_R +\bar\Delta_R$) and  
$x[(\mathbf{3,3,1},-2/3)+(\mathbf{{\bar 3},3,1},2/3)]\oplus 
y[(\mathbf{3,1,3},-2/3)+(\mathbf{{\bar 3},3,1},2/3)$
to have successful gauge coupling unification. If we use the Higgs spectrum from
Ref.\cite{okada} in the context of our model, we find that
$x\oplus{y}$ (from {\bf 210}) and $\Delta_L +
\bar{\Delta}_L \oplus \Delta_R +\bar{\Delta}_R$ are at about  the
$v_R$ scale.  Using these fields,  the Fig. \ref{Fig.3} shows
the coupling unification in this model and the gauge unification happens at
about $10^{15.5}$ GeV. We have chosen the $x,\,y$ and
$\Delta$ Higgs masses to be 6 times the $v_{R}$ scale to get this
value.  

First of all it is gratifying that the type I model is compatible with
gauge coupling unification with the desired value of $v_{R}$
scale. Furthermore the lowering of the ultimate unification scale implies
that the gauge exchange contributions to the proton decay in this model
lead to proton lifetimes much lower than the MSSM unifying without an
intermediate scale and will be around the current lower limit $5\times
10^{33}$ yrs when we
include the threshold corrections \cite{us}. The profile of the proton
decay modes in
the type I case will be substantially different from the pure type II case
recently discussed \cite{nasri} and is presently under investigation.

\section{Lepton Flavor Violating Processes and Lepton Edm}

\subsection{Lepton Flavor Violation}

We now discuss the lepton flavor violating processes e.g. $\mu\rightarrow
e\gamma$, $\tau\rightarrow
\mu\gamma$ etc. 
The operator for $l_i\rightarrow l_j+\gamma$ is:
\begin{equation}\label{eq501}
{\cal L}_{l_i\rightarrow l_j\gamma} = \frac{i e}{2 m_l} \, \overline{l_j}
\,\sigma^{\mu\nu} q_{\nu} \left( a_l P_L + a_r P_R \right) l_i
\cdot A_\mu + h.c.
\end{equation}
where $P_{L,R} \equiv (1 \mp \gamma_5)/2$ and $\sigma^{\mu\nu}
\equiv \textstyle{\frac{i}{2}} \,[ \gamma^{\mu}, \gamma^{\nu} ]$.
The decay width for $l_i\rightarrow l_j+\gamma$ can be written as:
\begin{equation}\label{eq502}
\Gamma(l_i\rightarrow l_j+\gamma) = \frac{m_\mu e^2}{64\pi} \left( |a_l|^2 +
|a_r|^2 \right)
\end{equation}
Then the branching ratio is obtained by multiplying this decay width with the
life time of the $l_i$ lepton.
The supersymmetric contributions include the neutralino and
chargino diagrams \cite{lfv}.

We work in the basis where the charged lepton masses are
diagonal at the highest scale of the theory. We first start with pure type II seesaw
model. We assume that $v_R$ 
(the scale where $G_{2231}$ breaks down to SM) is at the GUT scale. So
we have just MSSM and right handed neutrinos below the GUT scale. The right
handed masses have hierarchies and therefore get decoupled at different
scales. The flavor-violating pieces present
in $Y_{\nu}$ induces flavor violations into the charged lepton couplings and into
the soft SUSY breaking masses e.g.  $m^2$ terms etc. through the
following RGEs:
\begin{eqnarray}
dY_{e}/dt &=& {1 \over 16 \pi^2}(Y_\nu Y_\nu^{\dag}+\cdots)Y_e\\\nonumber
dm^{2}_{LL}/dt &=& {1 \over 16 \pi^2}(Y_\nu Y_\nu^{\dag}m^2_{LL}+
m^2_{LL}Y_\nu Y_\nu^{\dag}+\cdots)
\end{eqnarray}

\begin{figure}[tbp]
%    \centerline{ \DESepsf(muegmtyp2tan1040.eps width 8cm)}
    \centering
    \includegraphics*[angle=0,width=8cm]{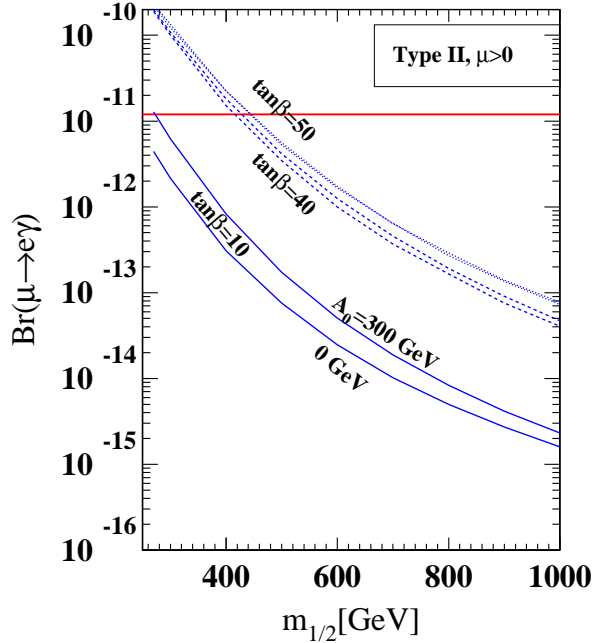}
    \vspace{0.1cm}
    \caption{\label{fig:fig1}  The BR[$\mu\rightarrow
e\gamma$] is plotted as a function of $m_{1/2}$ for different values
$A_0$ and $\tan\beta=10$, 40 and 50 in pure type II.}
\end{figure}

\begin{figure}
%    \centerline{ \DESepsf(muegmtyp1tan1040.eps width 8cm)}
    \centering
    \includegraphics*[angle=0,width=8cm]{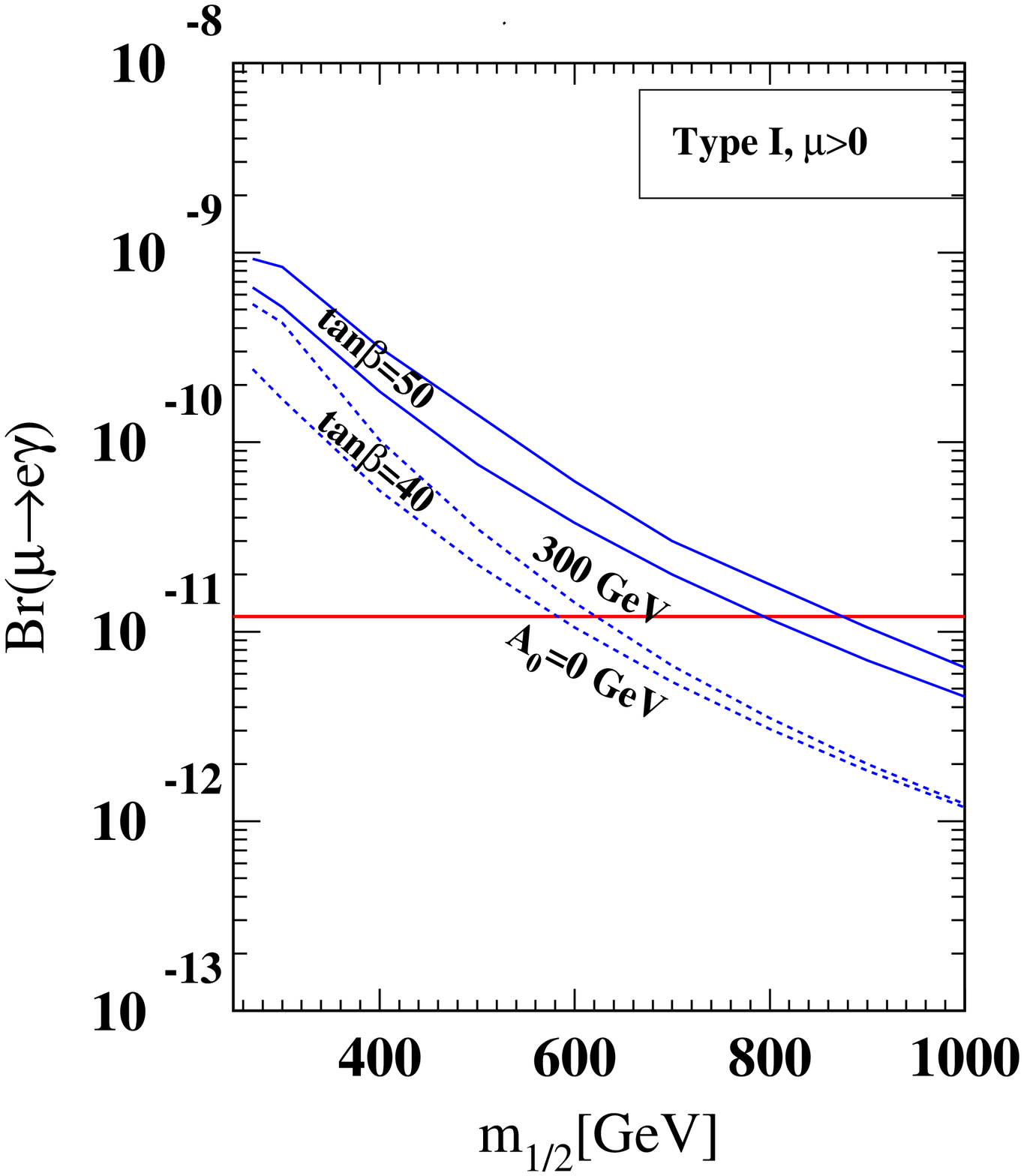}
    \vspace{0.1cm}
    \caption{\label{fig:fig2}  The BR[$\mu\rightarrow
e\gamma$] is plotted as a function of $m_{1/2}$ for different values $A_0$ and $\tan\beta=40$
and 50 in type I.}
\end{figure}

\begin{figure}
%    \centerline{ \DESepsf(taumugmtyp12.eps width 8cm)}
    \centering
    \includegraphics*[angle=0,width=8cm]{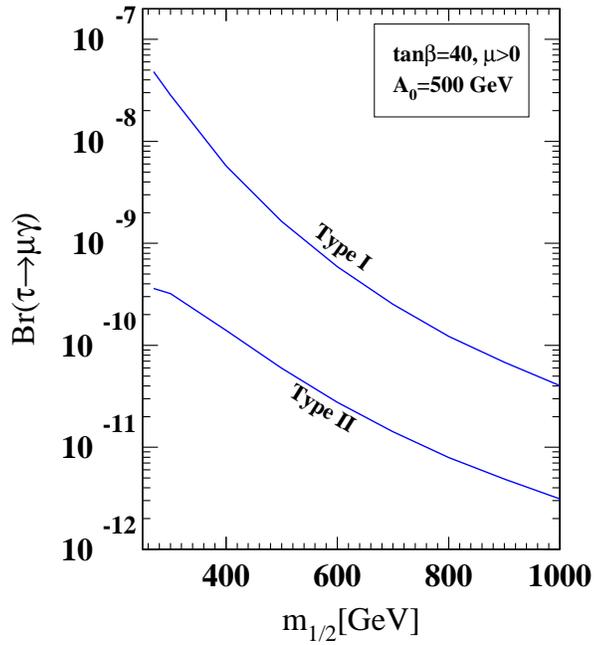}
    \vspace{0.1cm}
    \caption{\label{fig:fig3}  The BR[$\tau\rightarrow
\mu\gamma$] is plotted as a function of $m_{1/2}$ for 
 $\tan\beta=40$ in pure type II and  type I.}
\end{figure}

We use the mSUGRA universal boundary conditions at the GUT scale to draw
the figures. 
In Figs. \ref{fig:fig1}-\ref{fig:fig3}, we show the BR[$\mu\rightarrow
e\gamma$] and
BR[$\tau\rightarrow \mu\gamma$] as a function of $m_{1/2}$ for different values
of $A_0$. The lightest neutralino is the dark matter candidate in this model
and we satisfy the 2$\sigma$ range of the recent relic density constraint 
$\Omega_{\rm CDM}=0.1126^{+0.008}_{-0.009}
$\cite{wmap} in the parameter space. When we satisfy the relic density
constraint, the $m_0$ gets determined. For example, $m_0$ varies between 60-100
GeV for
$A_0=0$ GeV line in the graph. The Figs also show that the larger  
$\tan\beta$ has larger lepton flavor violation.

In type I model, the upper bound of $v_R$, VEV of $\overline{\mathbf{126}}$
which couples to fermions, is determined and this
value is not the GUT scale. 
For example, $v_R$ can be $\leq 10^{14}$ GeV.
% for $\tan\beta=10$ case. 
Then, 
$G_{2231}$ symmetry can be maintained between the GUT and the $v_R$ scale
if other Higgs fields do not break $B-L$ symmetry. 
This feature
induces larger lepton violation since the right handed neutrino is a part of the
doublet which has right handed electrons. The right slepton masses get new
flavor violating contributions through the flavor violating pieces present in
$Y_\nu$. The new contributions to the RGEs:
\begin{eqnarray}
dY_{e}/dt &=& {1 \over 16 \pi^2}(4 Y_\nu Y_\nu^{\dag}+\cdots)Y_e\\\nonumber
dm^{2}_{LL}/dt &=& {1 \over 16 \pi^2}(Y_\nu Y_\nu^{\dag}m^2_{LL}+
m^2_{LL}Y_\nu Y_\nu^{\dag}+2 Y_\nu m^2_{RR}Y_\nu^{\dag}+\cdots)\\\nonumber
dm^{2}_{RR}/dt &=& {1 \over 16 \pi^2}(Y_\nu Y_\nu^{\dag}m^2_{RR}+
m^2_{RR}Y_\nu Y_\nu^{\dag}+2 Y_\nu m^2_{LL}Y_\nu^{\dag} \nonumber \\
&&+ \frac32 [ff^{\dag}m^2_{RR}+m_{RR}^2 ff^{\dag}+2f m_{RR}^2 f^{\dag}]+\cdots) \nonumber
\end{eqnarray}
The new effects arising from the RGEs make the lepton flavor violations to be  
 larger in this case and is depicted in
Fig. \ref{fig:fig2}. 
In Fig. \ref{fig:fig3}, we plot $\tau\rightarrow \mu\gamma$ and find that
the Br can be as large as
$10^{-8}$ which can be explored in the near future. 

%\subsubsection{$\mu \rightarrow e \gamma$ and $\tau \rightarrow e \gamma$}

\subsection{Electric Dipole Moment of Electron}
\begin{figure}
%    \centerline{ \DESepsf(edmtyp2tan1040.eps width 8cm)}
    \centering
    \includegraphics*[angle=0,width=8cm]{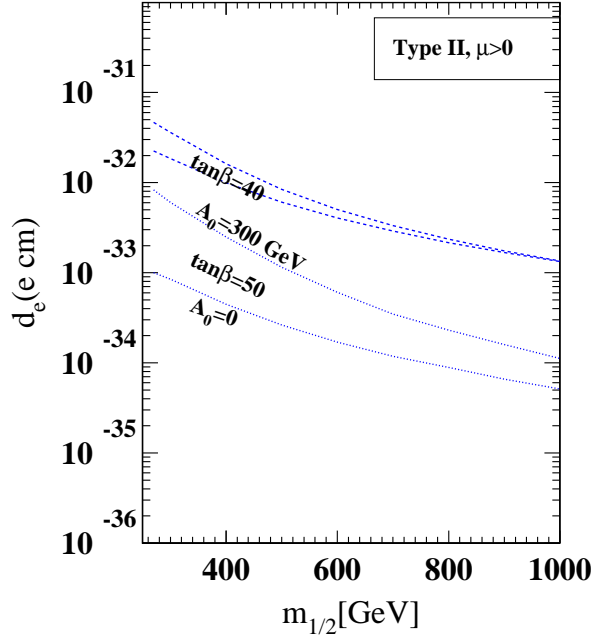}
    \vspace{0.1cm}
    \caption{\label{fig:fig4}  The electron EDM is plotted as a function of
    $m_{1/2}$ for different values $A_0$ and $\tan\beta=40$
 and 50 in pure type II.}
\end{figure}

\begin{figure}
%    \centerline{ \DESepsf(edmtyp1tan1040.eps width 8cm)}
    \centering
    \includegraphics*[angle=0,width=8cm]{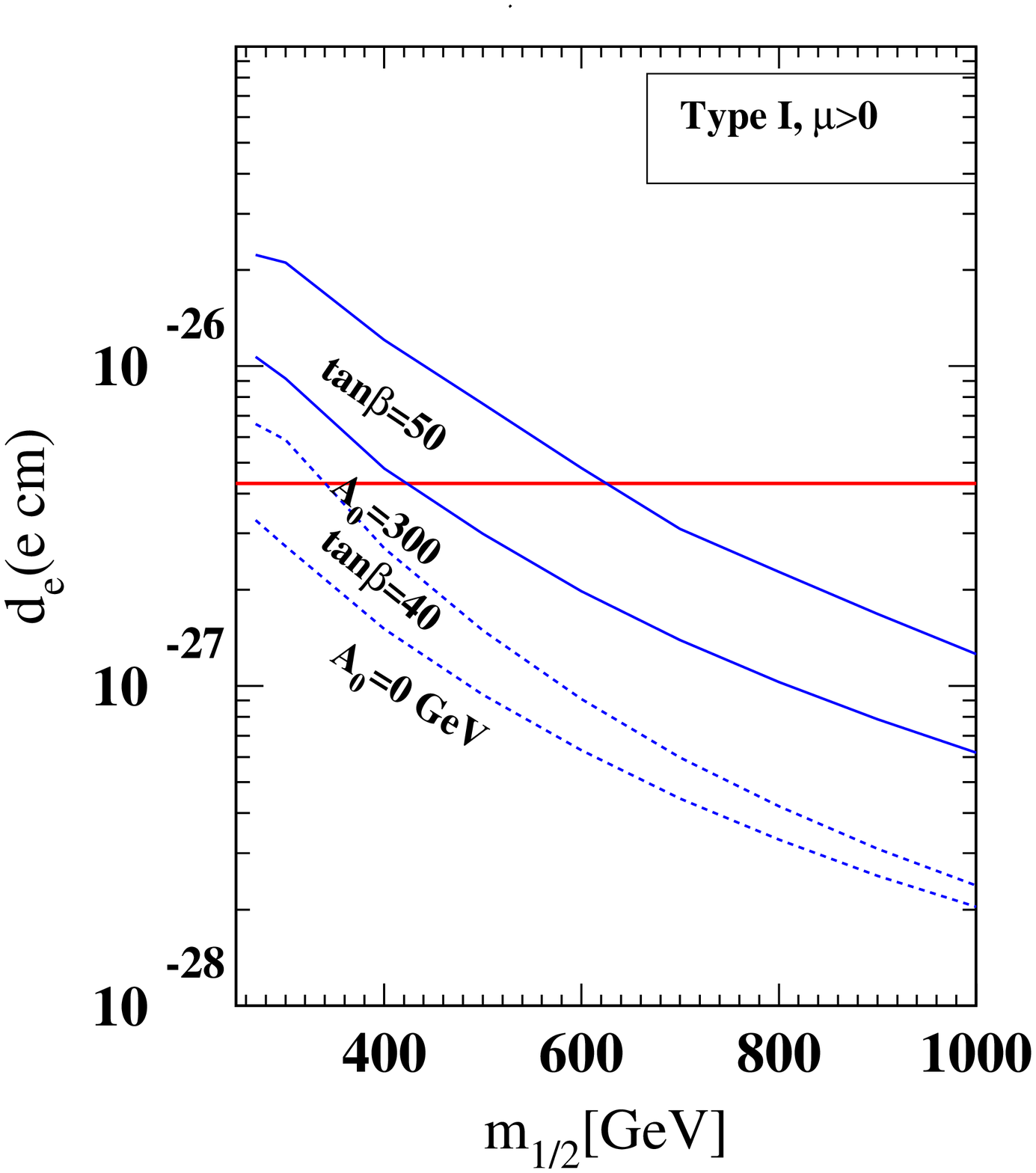}
    \vspace{0.1cm}
    \caption{\label{fig:fig5}  The electron EDM is plotted as a function of $m_{1/2}$ for different values $A_0$ and 
    $\tan\beta=40$
and 50 in  type I.}
\end{figure}

\begin{figure}
%    \centerline{ \DESepsf(edmtyp1tan1040.eps width 8cm)}
    \centering
    \includegraphics*[angle=0,width=8cm]{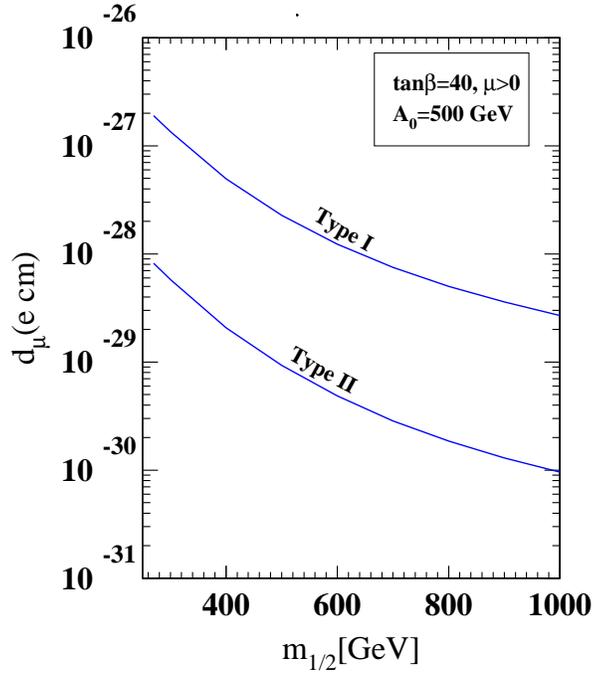}
    \vspace{0.1cm}
    \caption{\label{fig:fig6}  The muon EDM is plotted as a function of
$m_{1/2}$ is plotted as a function of $m_{1/2}$ for 
 40 in pure type II and  type I.}
\end{figure}

The EDM, $d_f$ for fermion $f_i$ appears in the effective Lagrangian as:
\begin{equation}
L_f={i\over 2}d_f\bar f\sigma_{\mu\nu}\gamma^5 fF^{\mu\nu}
\end{equation}
We have contributions from the chargino  and neutralino diagrams to
$d_f$\cite{df}. The
electron EDM is plotted in Fig. \ref{fig:fig4} for the type II and in
Fig. \ref{fig:fig5} 
for the type I. We again use the
same SUSY parameter space as in the case of lepton flavor violation. We
find that 
the  maximum value of EDM is $\sim
10^{-31}$ ecm for the type
II. For the type I, the EDM is large and is around $10^{-26}$ ecm and a large
region of parameter
space can be explored very soon. The muon EDM is shown in
Fig. \ref{fig:fig6} and the maximum value shown
is about $10^{-27}$ ecm and the scaling is broken in this model.

The $\sin2\beta$ calculated from the $B^0\rightarrow\phi K_s$ mode in
this model is 0.67, which
is not different from the SM prediction.  


\section{Conclusion}
     In summary, we have revisited the minimal renormalizable
SO(10) models with a {\bf 126} Higgs multiplet that has recently been
shown to predict neutrino mixings in agreement with observations, with
the primary goal of reconciling CKM CP violation with successful neutrino
predictions. We  consider
the most general type II and type I seesaw formula for neutrino masses
that includes the effect of the right handed neutrino mass matrix.
We find that in these models the basic ingredients that 
went into understanding maximal neutrino mixings i.e. the relation
$M_\nu~=~c(M_d-M_{e})$, can be recovered in
certain limits.  However they do not help to keep the
CKM phase in the first quadrant and in the type I case are incompatible with
detailed neutrino data. We remedy this 
by including a specific class of nonrenormalizable terms in the Yukawa
superpotential, that follows from a simple high scale theory. This specific set 
of nonrenormalizable terms also reconcile pure type II models with the
CKM model of CP violation.  We find
that the type I model requires  an intermediate $v_R$ scale
(i.e. $v_{R}\ll M_U$) which is lower than the standard GUT 
scale, however it is compatible with gauge coupling
unification which happens around $10^{15.5}$ GeV.  The proton life time 
is  around the current experimental limit. We also find that the mixed
type II and type I models require a value of the supersymmetry
parameter $\tan\beta$ larger than 30 to be compatible with present neutrino
data.   We then study the
phenomenological implications of
the type I, the mixed type II and the pure type II models for neutrino mixings, lepton
flavor violation and lepton edms.
We find that these predictions provide a new way to test these models.

\section*{Acknowledgments}

Y.M. thanks N.~Okada for usuful discussion
and KEK Theory Group for warm hospitality.
This work of B.D. and Y.M. is supported by 
the Natural Sciences and Engineering Research Council of Canada and 
the work of R. N. M. is supported by the National Science Foundation
 Grant No. PHY-0099544.

\end{document}